# Understanding the intrinsic compression in polycrystalline films through a mean-field atomistic model

Enrique Vasco[1*], María J. Ramírez-Peral[1,2], Enrique G. Michel[2,3,4] and Celia Polop[2,3,4]

[1]*Instituto de Ciencia de Materiales de Madrid, CSIC, Sor Juana Inés de la Cruz 3, 28049 Madrid, Spain*
[2]*Departamento de Física de la Materia Condensada, Universidad Autónoma de Madrid, 28049 Madrid, Spain*
[3]*Condensed Matter Physics Center (IFIMAC), Universidad Autónoma de Madrid, 28049 Madrid, Spain*
[4]*Instituto Universitario de Ciencia de Materiales Nicolás Cabrera, Univ. Autónoma de Madrid, Spain*

(Dated: September 30, 2020)

## ABSTRACT

Mullins' theory predicts the buildup of adatoms during surface diffusion at the edges of grooves where grain boundaries emerge to the surface of a polycrystalline film. However, the mesoscopic nature of this theory prevents the identification of the atomic-scale physical mechanisms involved in this phenomenon. Here, we interpret the buildup of adatoms in atomistic terms through a mean-field rate-equation model and demonstrate both its kinetic nature and its impact on the intrinsic stress in these systems. Furthermore, the model provides estimates of the surface profile of intrinsic stress, of its typical mean values, and of the dependence of stress on temperature and deposition flux for different growth stages. These estimates agree well with reported experimental results obtained from recent advances in nanoscale mapping of mechanical stresses on the surface of polycrystalline films.

# 1. INTRODUCTION

Compact polycrystalline films typically show a latent compression state upon completion of the film that is [1, 2, 3, 4]: (a) regenerable under conditions of high surface mobility, and (b) independent of film thickness. It could be described as a ticking time bomb for applications and devices requiring a long lifespan and/or non-mild operating conditions since regeneration and build-up of intrinsic stresses results in premature thermomechanical fatigue that underlies most mechanical failures. The origin of this behavior has been recently explained from a mesoscopic approach based on the Mullins´ theory [5,6]. It is due to the gradient of the surface chemical potential around the boundaries between the grains that form the polycrystalline solids, which causes the surface to evolve towards non-equilibrium morphologies. Among many other proposed explanations (reviewed elsewhere [7]), this stands out as the only one to date supported by direct measurements of mechanical stresses at the nanoscale [8], which is the characteristic spatial scale of stress in solids. There are only few atomistic studies on how diffusion around the grain boundary (GB) affects intrinsic stress in polycrystalline films (e.g., by kinetic Monte Carlo in Ref. [9]), and none of them addresses the recently proposed explanation.

In this work, we develop an atomistic mean-field model that sheds light on the physical processes involved in the mesoscopically described phenomenon [5,6]. In particular, we address the origin of the surface profiles around the GB, which exhibit out-of-equilibrium shapes [8] characterized by the accumulation of mass (buildup of adatoms) at the edges of the groove (or gap [10]) where the GBs emerge to the surface. The kinetic nature of the phenomenon is probed and used to estimate the resulting profile of intrinsic stress under conditions of high atomic mobility. This description allows us to explain the dependence of the intrinsic stress on the deposition parameters, in order to address the large body of experimental evidence reported so far.

## 1.1. Review of the mesoscopic Mullins´ theory

In this study, we assume a surface, which is initially in equilibrium, evolving under deposition conditions. As previously discussed in Refs. [5,6], the equilibrium profile $h_e$ in the vicinity of a GB counters the surface gradient of the balance $\vec{V}_S \gamma$ between interfacial tensions along the GB groove profile (from the GB triple-junction to far-from-the-GB free surface). This causes the stress normal to the surface (i.e. $//\widehat{N}$) $\vec{\sigma}_N = -(\vec{V}_S \cdot \gamma)\widehat{N}$, as described by the Young-Laplace equation, to be cancelled. The equilibrium profile $h_e(r)$, with $r$ denoting the surface position, can be approximated by the expression [11]:

$$h_e(r) = 1.77 h_{GB} ierfc[m_0(r - r_{GB})/1.77 h_{GB}] \qquad (1),$$

where $h_{GB}$ is the GB triple-junction depth measured from the average height of the surface $\langle h \rangle$ (which corresponds to the film thickness), $ierfc(x)$ denotes the integrate of the complementary error function, $m_0 = tan\{arcsin[\gamma_{GB}/2\gamma_s]\}$ is the equilibrium surface slope at the GB triple-junction and $r_{GB}$, the GB position ($r_{GB}=0$ hereafter). The nature of the equilibrium profile $h_e(r)$ in Eq. 1 is discussed in Refs. [12,13,14].

Under deposition conditions far from equilibrium, the surface profile $h(r,t)$ evolves according to the mesoscopic Mullins´s equation [11,14]

$$\frac{1}{\Omega}\partial_t[h(r,t) - \langle h \rangle] = \vec{V}_S \cdot \vec{J} \propto -\nabla^4 h(r,t) \qquad (2)$$

driven by a surface diffusion current $\vec{J} = -\phi \vec{\nabla}_s \kappa$. $\vec{J}$ is biased by the gradient of the surface curvature $\vec{\nabla}_s \kappa$ (driving force) where $\kappa \propto \nabla^2 h$; $\phi$, the diffusive mobility of adatoms, corresponds to the kinetic constant of $\vec{J}$; and the minus sign considers the curvature sign convention [6]. Surface curvature provides an estimate of the local surface chemical potential in terms of relative density of dangling bonds $n_{db} \propto \kappa$. Thus, $\vec{J}$ flows from the concave regions (where the density of dangling bonds is lower) towards convex ones (with higher density) through the flat areas.

Considering the surface slope constraints at the GB triple-junction and away from GB (namely, $\partial_r h(r,t)|_{r \to r_{GB}} \to m_0$ and $\partial_r h(r,t)|_{r \to \infty} \to 0$, respectively) (see Fig. 6 of Ref. [6]), the solution of the Mullins equation causes the initial equilibrium profile to evolve into a kinetic state. This state is characterized by the mass accumulation at the edges on both sides of the GB groove [5,6,11]. In atomistic terms: (a) this mass accumulation can be understood as the buildup of adatoms with volume $\Omega$, while (b) the slope constraint models the existence of local gradients in the density of steps between surface terraces. The physical mechanisms underlying mass accumulation are difficult to identify from the numerical solution of the mesoscopic model that involves a partial fourth-order differential equation (Eq. 2). This fact, together with experimental difficulties in detecting mass accumulations in steep regions near the GB groove by Atomic Force Microscopy (AFM), does not allow us to rule out the existence of mathematical artifacts a priori.

We infer the following from the mesoscopic model, [5,6]: The continuity equation (left part of Eq. 2) predicts that the accumulation must be induced by a non-null surface divergence $\vec{\nabla}_S$ of $\vec{J}$, which can be attributed to local changes in both the diffusive mobility $\phi$ of the adatoms (i.e., $\vec{\nabla}_S \phi \neq 0$) and/or the diffusion driving force [$\vec{\nabla}_S \cdot (\vec{\nabla}_S \kappa) \neq 0$] along the profile of the GB groove:

$$\vec{\nabla}_S \cdot \vec{J} = -\vec{\nabla}_S \phi \cdot \vec{\nabla}_S k - \phi \vec{\nabla}_S \cdot (\vec{\nabla}_S k) \qquad (3).$$

In the case of the Mullins-type diffusive profile, the mass accumulation at the edges of the GB grooves has been ascribed to the decrease in the diffusive mobility of the adatoms as they diffuse towards the GB triple-junction. Plausibly and in agreement with the results in [9], diffusive mobility falls within the groove as the density of step-edge barriers to cross increases. This effect would reduce interlayer transport, forcing the diffusing adatoms to meet and nucleate reversibly. On the other hand, although the surface curvature "diverges" at the GB triple-junction, causing the density of dangling bonds to reach a long-range maximum, the constriction of the surface slope limits the transport throughput (see the effective local rate of surface advance in Fig. 6 of Ref. [6]). This causes the gradient of superficial curvature to remain roughly constant ($\vec{\nabla}_S k \approx const.$) along the groove profile such that its divergence is minimized. In practical terms, the second term in Eq. 3 is much smaller than the first, and thus $\vec{\nabla}_S \cdot \vec{J} \propto -\vec{\nabla}_S \phi$.

In this context, a microscopic interpretation of the profile of the GB groove in terms of terraces and steps (vicinal surfaces), rather than the blind distribution of dangling bonds, helps us to identify the physical mechanisms involved in the phenomenon of mass accumulation induced by the Mullins-type diffusion on the surface of polycrystalline solids. As we know today [5,6,8], this phenomenon underlies the origin of the intrinsic stress experienced by these systems during their lifespan.

## 2. MODEL, RESULTS AND DISCUSSION

### 2.1. Atomistic approach to growth by step-flow

By applying the law of conservation of mass, the evolution of the surface profile (in Eq. 2) can be rewritten in atomistic parameters as:

$$\frac{1}{\Omega}\partial_t h(r,t) = F - \partial_t n_1(r,t) \quad (4),$$

where $n_1(r,t)$ is the density of diffusing monomers (i.e. mobile single adatoms), which are deposited from a flux $F$ and incorporated into the film bulk at the growth rate. Thus, Eq. 4 distinguishes between the deposition rate $F$ (atoms deposited per unit of time and area) and the growth rate $\partial_t h$ (thickness variation over the unit of time) through $n_1(r,t)$ evolution, which is described according the following kinetic rate equation:

$$\partial_t n_1 = F - D_s n_1 n_{step} - D_s[1 - (\lambda/\xi)^2]n_1^2 + \beta D_s \nabla_s^2 n_1 \quad (5)$$

where $D_s$ is the intralayer monomer diffusion coefficient, $n_{step} = \frac{3}{2}\left(\sqrt{2}/\partial_r h(r,t) - 3/4\right)^{-1}$ denotes the step density in a vicinal surface comprising fcc (111) terraces separated by indistinguishable A and B steps [15], $\lambda/\xi$ corresponds to the diffusion length-to-grain size ratio, and $\beta$ is a kinetic parameter described below. Since metal (111) planes with the highest close-packing and lowest energy cause most polycrystalline metal films to grow [111]-textured with high roughness (following a Volmer-Weber mode), it makes sense to use a (111) vicinal surface as a generic growth scenario.

Eq. 5 corresponds to the easiest form of kinetic rate equation able to consider the following processes: The first term takes into consideration that the flux feeds evenly the monomer density; the second and third terms account for the decay in $n_1$ by monomer capture by steps and monomer capture by other monomers, respectively. The fourth term considers the monomer diffusion driven by surface gradients of $n_1$, which corresponds to the thermodynamic potential involved in the Fick's first law $\vec{j}_1 = -D_s \vec{\nabla} n_1$. The ability of the fourth term to homogenize spatially the monomer density is controlled by $\beta$. Eq. 5 assumes low growth temperatures, such that the re-evaporation is negligible. Figure 1c offers a visual description of the atomistic parameters involved in the model.

Fig. 1a displays the equilibrium profile (black curve) in the vicinity of a GB with $m_0$ = 0.3 [16], along with its densities of (111) steps ($n_{step}$, red curve) and dangling bonds according to Mullins´s equation ($n_{db} \propto \nabla^2 h$, dashed green curve). As revealed from the $n_{step}$ profile, the steps are concentrated within the GB groove. $n_{step}$ remains nearly constant close to the GB triple-junction where the surface slope is constrained to $m_0$ (gray line) and decreases towards the grain surface. Note that the $n_{step}$ profile differs greatly from the $n_{db}$ profile, as considered by the Mullins´ theory [11] that predicts an accumulation of dangling bonds at the GB triple-junction where the surface curvature diverges. Since in the polycrystalline films, the intrinsic stress relaxes (down to residual levels of two-three orders of magnitude lower) under low surface mobility conditions (e.g., at room conditions), the initial surface profiles (at $t = 0$) shown below are taken to be similar to $h_{GB}$-normalized equilibrium profiles [i.e., $h(r, t = 0) \equiv h_e(r)/h_{GB}$]. This is because major differences between initial and equilibrium profiles would imply high residual levels of stress (see discussed in section 2.2).

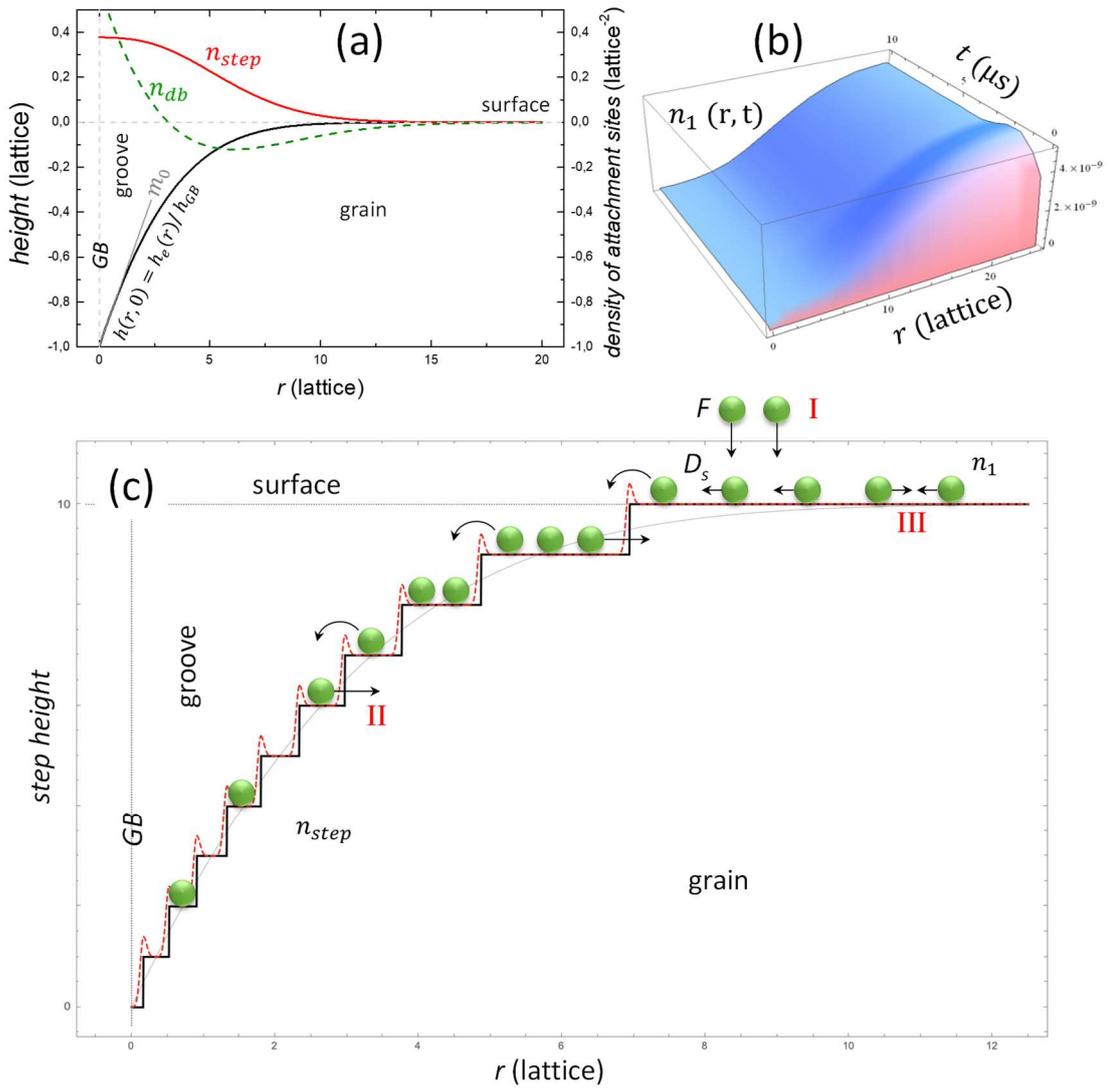

**Fig. 1 (a)** $h_{GB}$—normalized equilibrium surface profiles [$h(0) = h_e$, black curve] along with the density of A/B-undistinguishable (111) steps ($n_{step}$, red curve) and dangling bonds ($n_{db}$, dashed green curve) around a GB triple junction at $r_{GB}$=0 (with $m_0$=0.3 [16], gray line). **(b)** Evolution of the profile of the monomer density computed from Eq. 5 with $F$ =0.5 nm/min (≈2 ML/min), representative $D_s$=100 μm²/s at RT [17] (≈1.6×10⁹ lattice²/s for Au(111) with lattice parameter $a$ =0.252 nm), $\beta = 0.01$ and $\xi = \lambda$ (i.e., kinetically limited growth by step-flow without second nucleation). **(C)** Schematic atomistic landscape of the model. Roman numbers denote the terms in Eq. 5: (I) deposition rate, (II) step-flow and (III) second nucleation. The dashed red curve traces the potential-energy profile of the surface including step-edge barriers.

By inserting Eq. 5 in Eq. 4, the local rate of surface advance is described as follows:

$$\frac{1}{\Omega}\partial_t h = D_s n_1 n_{step} + D_s[1 - (\lambda/\xi)^2]n_1^2 - \beta D_s \nabla_s^2 n_1 \qquad (6).$$

The first term in Eq. 6 corresponds to growth by step-flow within the GB groove (i.e., steps capture diffusing monomers). The second term accounts for second nucleation on the grain top (diffusing monomers meet each other), and the third one correlates spatially the rates of the two first terms through the monomer density.

The factor $[1 - (\lambda/\xi)^2]$ in the second term evaluates the probability of second nucleation on the grain top. For grain sizes much larger than the diffusion length ($\xi \gg \lambda$), the contribution of the monomers to the step-flow is proportional to the number of atoms landing in the $\lambda$-wide strip near the GB (where $n_{step}$ is higher), and then the monomer current is $j_1 \sim F\lambda(\lambda/\xi)$. Otherwise, all the deposited monomers contribute with $j_1 \sim F\xi$. In accordance with the law of conservation of mass, the probability of second nucleation can be estimated from the relative difference between these two cases $\propto [F\xi - F\lambda(\lambda/\xi)]/F\xi$. In principle, the diffusion length is effectively limited by the grain size, so that $[1 - (\lambda/\xi)^2]$ is ranged between 0 and 1 (i.e., without and with second nucleation, respectively).

On the other hand, the kinetic parameter $\beta$ (in the third term of Eq. 6) ranges between two limiting cases: $\beta \approx 1$ corresponds to a kinetic limitation-free stage where the surface diffusion is much faster than the monomer capture and cancels any gradient of the monomer density. Otherwise, $\beta \ll 1$ corresponds to an aggregation-limited diffusion that characterizes kinetically limited growth. The interlayer transport delayed by step-edge barriers and/or reversible nucleation would be, potentially, the major kinetic limitations in these systems.

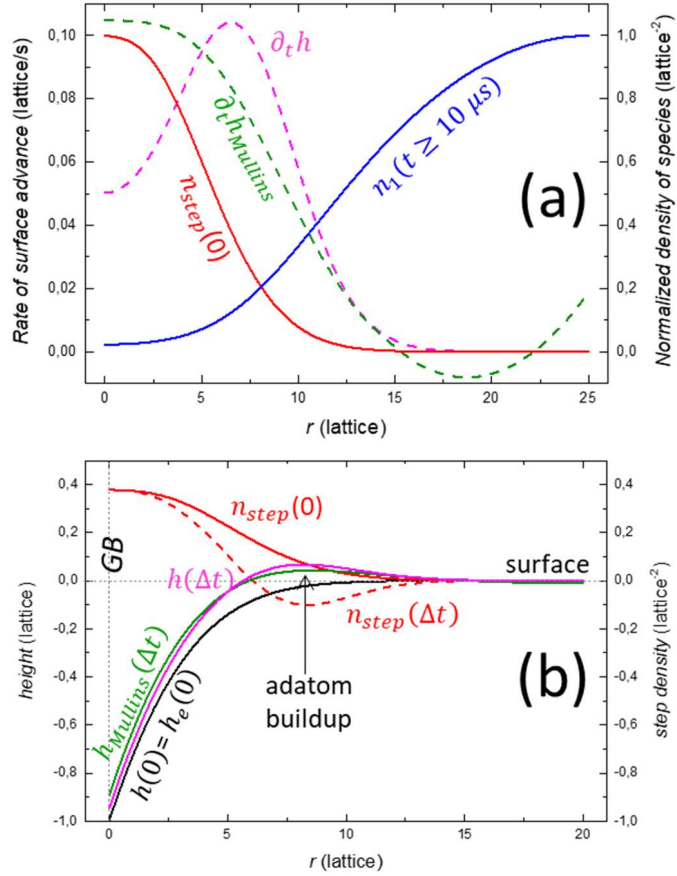

**Fig. 2 (a)** Profiles of: the steady monomer density ($n_1$, blue curve), the density of A/B-undistinguishable (111) steps ($n_{step}$, red curve), and the local rates of surface advance by step-flow ($\partial_t h$, dashed pink curve) and according to Mullins's equation ($\partial_t h_{Mullins}$, dashed green curve) around a GB triple junction at $r = 0$ (with $m_0$=0.3, $\beta$=0.01 and $\xi = \lambda$). **(b)** Evolution of the surface profile (only one side is shown) after $\Delta t$ =1 s [21] growth by step-flow [compare the final profile (pink curve) with the numerical solution of Mullins's equation (green curve), both obtained from the initial profile (black curve)]. The profiles of the densities of (111) steps before and after 1 s-growth (solid and dashed red curves, respectively) are included.

Fig. 1b shows the evolution of the $n_1$ profile computed by integrating Eq. 5 for a kinetically limited growth by step-flow (i.e., with $\beta = 0.01$ and $\xi = \lambda$). $n_1$ profile (along x-axis) stabilizes to steady values in fractions of μs (y-axis), which are different inside ($n_1 \to F/D_s n_{step}$) and outside ($n_1 \to F\left(\frac{\xi^2}{2}\right)/D_s$) the GB groove. These values differ in several orders of magnitude (~two orders in our study for $\xi/2 = 25$ lattices), which indicates that monomer density is depleted within the GB groove, in agreement with the predictions of the classical growth theories [18]. This result rules out previous models that attributed the postcoalescence compression to the insertion/trapping of adatoms in GBs [4,19,20] because since the GB groove is empty in monomers there is nothing to insert or capture. In this context, it is also interesting to note that Mullins´s theory predicted early that currents involving transport along GBs have a negligible effect on the material relaxation (see Eq. 15 in Ref. [11]).

(**Kinetically limited growth, $\beta \ll 1$**) Fig. 2a shows the local rate of surface advance by step-flow (first term in Eq. 6, $\partial_t h$ —dashed pink curve). The normalized profiles of densities of the involved species in this type of growth, namely, saturated $n_1$ (blue curve, for $t \geq 10$ μs according to Fig. 1b) and initial $n_{step}$ (red curve taken from Fig. 1a) are also included. Since both densities have opposite behaviors ($n_1$ rises as we move away from the GB while $n_{step}$ drops), $\partial_t h$ exhibits a maximum where the product $n_1 \times n_{step}$ is highest. In other word, $\partial_t h$ is higher at the edge of the GB groove where both species coexist in moderate amounts. By way of comparison, the local rate of surface advance obtained from the slope-constrained Mullins´s equation ($\partial_t h_{Mullins}$ —dashed green curve taking from Fig. 6 of Ref. [6]) is also included in Fig. 2a. $\partial_t h_{Mullins}$ predicts higher growth within the groove, while $\partial_t h$ points to further accumulation on the edge. The negative values of $\partial_t h_{Mullins}$ beyond the edge mean a depletion of the grain at long term, because unlike $\partial_t h$, $\partial_t h_{Mullins}$ does not correspond to an initial rate of surface advance.

Fig. 2b shows the surface profile that results from adding the local rates of surface advance ($\partial_t h$ and $\partial_t h_{Mullins}$, in Fig. 2a) for a given step time $\Delta t$ (here $\Delta t$ =1 s [21]) to the initial profile [$h(0)$ —black curve] as: $h(r, \Delta t) \approx h(r, 0) + \Delta t \partial_t h(r, 0)$. The resulting microscopic profile [$h(\Delta t)$ —pink curve] shows a good agreement with the slope-constrained Mullins´s profile [$h(\Delta t)_{Mullins}$—green curve]. The accumulation of material at the edge of the GB groove, giving rise to ridges, is clearly visible in both profiles (see the arrowed volume), although this is smoother in the Mullins´s one.

The red curves display the evolution of $n_{step}$ profile before (solid) and after (dashed) the surface advance. The accumulation of material at the edge results from the meet and nucleation of the adatoms along the diffusion path towards the GB triple junction. This gives rise to the formation of small terraces with unstable steps where the monomer capture is reversible. Consequently, $n_{step}$ takes negative values to indicate that these sites correspond to delayed detachment sites (i.e., kinetic limitations as discussed above) rather than attachment sites.

(***Kinetic limitation-free growth, β → 1***) Fig. 3a displays the $\beta$—dependence of $n_1$ for a step-flow regime. As $\beta$ rises, the influence of the spatial distribution of the step density on the diffusion kinetics of the monomers decreases. The quick diffusion homogenizes spatially $n_1$ profile (along x-axis) in few µs (y-axis), as displayed in Figs. 3b. In the absence of major kinetic limitations (e.g. $\beta$=0.9), growth in regions with higher $n_{step}$ prevails independently on the site positions/proximity. Consequently, the local rate of the surface advance ($\partial_t h$ —dashed pink curve in Fig. 3c) reproduces well the shape of the normalized $n_{step}$ profile (red curve in Fig. 3c). Note that both curves (dashed pink and red) exhibit a high degree of overlap.

Fig. 3d shows the surface profile [$h(\Delta t)$ —pink curve] that results from the initial profile [$h(0)$ —black curve] considering the advance rate plotted in Fig. 3c after $\Delta t$ =1 s [21] of kinetic limitation-free growth by step-flow. The procedure is as the one followed to obtain the data plotted in Fig. 2b. However, the profile $h(\Delta t)$ that results in this case does not exhibit mass accumulation at the edge of the GB groove. On the contrary, a comparison with the equilibrium profile [$h_e(\Delta t)$ —dark-cyan curve] corresponding to a similar deposited volume [22] reveals a good agreement between both. This agreement indicates that kinetic limitation-free growth does not modify the equilibrium condition between interfacial tensions (described by Laplace-Young equation) as expected.

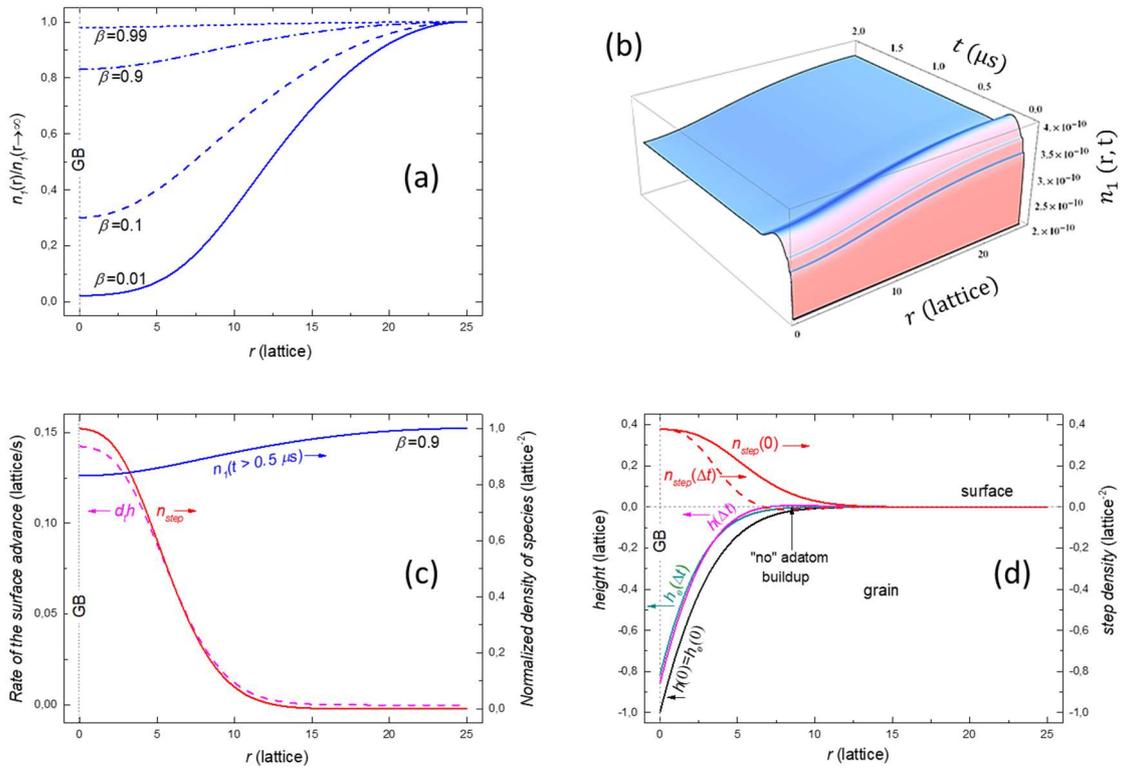

**Fig. 3 (a)** $\beta$—dependence of the $n_1$ profile around a GB triple junction at $r=0$ (with $m_0$=0.3 and $\xi = \lambda$) for growth by step-flow. $n_1$ profiles are normalized by their far-from-GB values (i.e., on the gain top). **(b, c and d)** Show similar results to those displayed in Fig. 1b, 2a and 2b, respectively, for the same set of parameters, except $\beta = 0.9$. Dark-cyan curve in (d) corresponds to the surface equilibrium profile for a similar deposited volume.

At this point, we can provide a microscopic description of the surface kinetics underlying the phenomenon of mass accumulation at the edges of the GB grooves, which addresses the results of the mesoscopic model [6]. The adatoms diffuse from the grain tops towards the GB grooves where the density of dangling bonds is higher (this is the thermodynamic driving force of the process). However, their diffusive mobilities (that determinate the kinetics of the process, i.e., its rate) are limited by the reversible aggregation to the closer steps (where $n_{step}$ is negative in Fig. 2b). In other words, while the long-range gradient of the density of steps biases the flow of diffusing adatoms, the short-range gradients control the flow rate and, consequently, the short-term surface advance.

## 2.2. Intrinsic Stress

Unlike the extrinsic stress in thin films, which is associated to lattice and thermal mismatches with the substrate, the intrinsic stress is attributed to unbalanced force fields that arise around discontinuities in the crystalline lattice of the films. In polycrystalline films, the lattices discontinuities playing major role in the generation of intrinsic stress are the grain boundaries and the film surface. As discussed above, the Laplace-Young equation describes the balance between these two discontinuities at the GB groove, with the Laplace pressure being the resulting stress for small perturbations in the equilibrium profile (Eq. 1). These perturbations must be understood in terms of changes in the chemical potential of the surface, which is estimated from its curvature $\kappa$. Consequently, an intrinsic stress $\sigma_N$ in the form of a Laplace pressure is generated from the Young-Laplace equation:

$$\sigma_N(r,t)/\gamma_s = 2[\kappa_k(r,t) - \kappa_e(r,t)] \tag{7}$$

whenever the curvature $\kappa_k$ of the surface profile that results from kinetically limited growth differs from the curvature $\kappa_e$ of the surface profile that corresponds to kinetic limitation-free growth (that of equilibrium as demonstrated above), for a similar deposited volume.

Fig. 4 shows the surface profiles generated after $\Delta t$=1 s-deposition under kinetically limited and not limited conditions (pink and green curves, respectively), whose curvatures (dashed curves) differ from each other. The orange curve plots this difference $(\kappa_k - \kappa_e)$, which according Eq. 7, provides an estimate of the local ratio of the normal stress $\sigma_N$ to the surface tension $\gamma_s$. For textured vicinal surfaces, formed by terraces with a preferential crystalline orientation, $\gamma_s$ is roughly constant and then $\sigma_N(r,t) \propto (\kappa_k - \kappa_e)$. The thus-estimated stress profile exhibits an oscillatory behavior with a prevailing compressive contribution at the edge of GB groove (here around $r \approx 7.5$ lattice from GB triple junction), which gives rise to a main compression averaged over the surface profile. From the data in Fig. 4, $\langle \sigma_N \rangle \approx 2\langle \kappa_k - \kappa_e \rangle \gamma_s \approx \left(\frac{2}{96.5}\right) \gamma_s/a \sim 2\%$ of $\gamma_s/a$, with $a$ denoting the lattice parameter. The typical features (oscillating behavior, spatial range and local magnitude) of this stress profile agree reasonably with those measured experimentally in Au and Cu polycrystalline films by Atomic Force Microscopy-related techniques [8]. Besides, the estimate of the main compression in Fig. 4 from Au(111) parameters ($\gamma_s \approx 1.54$ J/m² and $a = 0.252$ nm [23]) provides $\langle \sigma_N \rangle \approx 123$ MPa, which is within typical range of intrinsic stress jump reported for evaporated noble metals [20,24]. These results support the predictions of the models [3,5,6,25] that attribute the generation of intrinsic stresses to the emerging of non-equilibrium structural shapes and profiles due to surface kinetics.

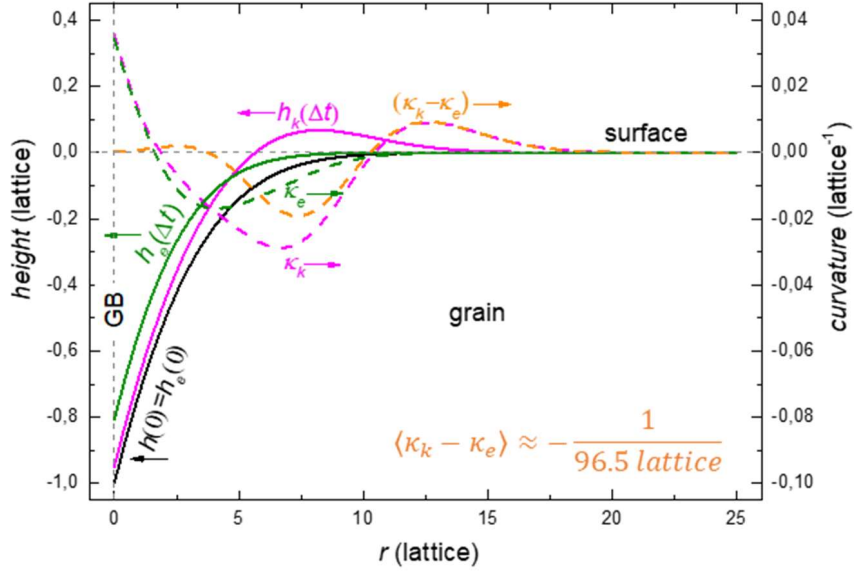

**Fig. 4.** Surface profiles (taken from Figs. 2b and 3d) resulting from kinetically limited (pink curve for $\beta$=0.01) and kinetic limitation-free (green curve for $\beta$=0.9) growths for the same deposited volume, together with their curvatures $\kappa_k$ and $\kappa_e$ (dashed curves of the same color), respectively. The difference between the curvatures $(\kappa_k - \kappa_e)$ is included (orange curve), together with its main value averaged over the surface profile.

## 2.3. Atomistic approach to growth by second nucleation

Since the diffusion length $\lambda$ is determined by the diffusion coefficient-to-flux ratio, as $\lambda^6 \sim D_s/F$ for 2D+1 growth of films from diffusing monomers [14,18], the variation of the flux $F$ and/or the deposition temperature $T$ gives rise to diverse effects at different growth stages. During early growth stages of homogeneous nucleation and coalescence on substrates that do not prompt long-range order (e.g. substrates with amorphous surfaces or highly incommensurate ones), $F$ (and complementary $T$) determines the average size of the grains $\xi \approx \lambda(\langle h \rangle \to 0)$, with $\langle h \rangle$ denoting the film thickness. Otherwise, the *in-situ* variations of $F$ and $T$ (later growth parameters) during the postcoalescence modify the fraction of the grain surface $\lambda(\langle h \rangle \gg 0)/\xi$ involved in the phenomena of mass transport. As the flux increases and/or the temperature drops, the transport towards the GBs decreases and the step-flow leaves room for the second nucleation on the grain top for $\lambda \ll \xi$. The second nucleation on the same grain, which forces the crystalline coherence of the second nuclei, gives rise crystallites with slight misorientations to each other (in agreement with the Structure Zone Model´s predictions for zone 1b [26,27]), which coalesce without generating new grain boundaries. Instead, low-angle and CSL defects are formed. This behavior is similar to the case of seeded substrates. In short, changes in the later growth parameters (flux and/or temperature) induce extra roughness on the surface of the grains without significantly modifying their sizes. Consequently, later $F$ (or $T$) and the average grain size (or the GB density) can be treated as independent growth parameters.

Based on the above considerations, the behavior of $\sigma_N$ with the early growth parameters can be addressed straightforwardly from the dependence of the GB density on these parameters [28]. Thus for example, higher $T_{early}$ (lower $F_{early}$) gives rise to larger grains, a lower density of GB sites between them, and consequently lower post-coalescence intrinsic compression. Conversely, the behavior of $\sigma_N$ with the later growth parameters is more difficult to understand. On the other hand, dependence on intrinsic stress with flux characteristics that do not substantially alter surface diffusivity (e.g. its spatial inhomogeneity leading to shadowing, steering…) are not considered here. This is because intrinsic stress generation has also been observed in high surface mobility conditions that do not involve flux (e.g. by post-deposition annealing [16] and bombardment with energy particles [28]).

Figure 5a shows the $\lambda/\xi$—dependences of the densities of the surface species (monomers and steps) involved in film growth by (i) step-flow (with local rate of surface advance $\partial_t h \propto D_s n_1 n_{step}$ at the edge of grain-boundary), and (ii) second nucleation (with $\partial_t h \propto D_s [1 - (\lambda/\xi)^2] n_1^2$ on the grain top). As $\lambda/\xi$ decreases (from solid curve to increasingly dashed ones, according to arrows directions), the second nucleation prevails depleting the grain top of monomers. This is because the second nucleation implies the meet of several monomers, such that $n_1$ decays at a rate of $\partial_t n_1 \propto -n_1^2$ down to a steady value of $n_1 \to \sqrt{F/D_s}$. Fig. 5b shows the resulting surface profiles ($h(r,t)$ —pink curves, which are calculated in a similar way to those displayed in Figs. 2b and 3d), together with the corresponding equilibrium profiles [$h_e(r,t)$ —green curves]. Such equilibrium profiles are estimated by assuming: (a) a similar deposited volume [22], and (b) a similar height far from the grain boundary [i.e. $h_k(r \to \infty, t) \equiv h_e(r \to \infty, t)$]. The curvature curves (like those displayed in Fig. 4) are not included in Fig. 5b for the sake of clarity. Instead, their differences are plotted [$(\kappa_k - \kappa_e)$ —orange curves]. As $\lambda/\xi$ decreases, the mass buildup on the edge of the grain-boundary groove is transferred to the grain top. This causes (not shown here—use Fig. 4 to follow the description) the minimum of curvature $\kappa_k$ of the resulting surface profile to shift towards the GB position, while the minimum of $\kappa_e$ is moving in the opposite direction. Thus, the curvature minima approach to each other, such that the overlap of the curvature curves results in a decrease in the minimum of their difference $(\kappa_k - \kappa_e)$, as shown in Fig. 5b. According to Eq. 7, this implies a decrease in both the compression maximum $max(\sigma_N)$ and its main value $\langle \sigma_N \rangle$.

Beyond the reported qualitative behavior of the intrinsic stress with the later deposition flux (namely, $\sigma_N$ drops as $F_{later}$ increases [29,30,31]), in this study, we aim to address the quantitative dependence between these two magnitudes. Fig. 5c shows the $\lambda/\xi$—dependencies of $max(\sigma_N)/2\gamma_s$ and $\langle \sigma_N \rangle/2\gamma_s$ (red and black symbols, respectively), which fit well to parabolic functions (dashed curves). Assuming that the behavior of these statistical values is representative of the intrinsic stress, and taking into account $\lambda^6 \sim D_s/F$, we get:

$$\sigma_N \sim (\lambda/\xi)^2 \sim \left(F_{early}/F_{later}\right)^{2/6} \propto F_{later}^{-1/3} \qquad (8),$$

where, as discussed above, $F_{early} = F(\langle h \rangle \to 0)$ is the early-growth flux, determining the grain size $\xi$. The power-law dependence $\sigma_N \propto F_{later}^{\alpha}$ (with $\alpha = -1/3$) in Eq. 8 is consistent with the later flux-dependences of the steady intrinsic stress $\sigma_\infty$ reported for moderated fluxes and submicron-sized grains [29,30,31], whose power-law exponents are ranged between $\alpha = -0.21$ and $-0.40$ as displayed in Fig. 5d.

Eq. 8 describes the effects of the flux on intrinsic stress for different stages of growth. Thus, while the early increase in flux raises the density of grain boundaries around which the material under compression accumulates (i.e. $\sigma_N$ increases), the late postcoalescence increase in flux induces extra surface roughness that results in traction between crystalline-coherent surface features (i.e. $\sigma_N$ decreases [29]). However, the crossover between both regimes changes for 3D growths by Volmer-Weber mode owing to the kinetic roughening of the surface as the film grows [14]. Hence, early flux is not low enough to avoid second nucleation throughout deposition, and consequently late increases in flux are not necessary to induce extra-roughness and traction on grain surfaces. This effect may consistently explain the non-monotonic behavior of $\sigma_N$ vs. flux (with a maximum of compression for intermediate fluxes) reported in Ref. [28].

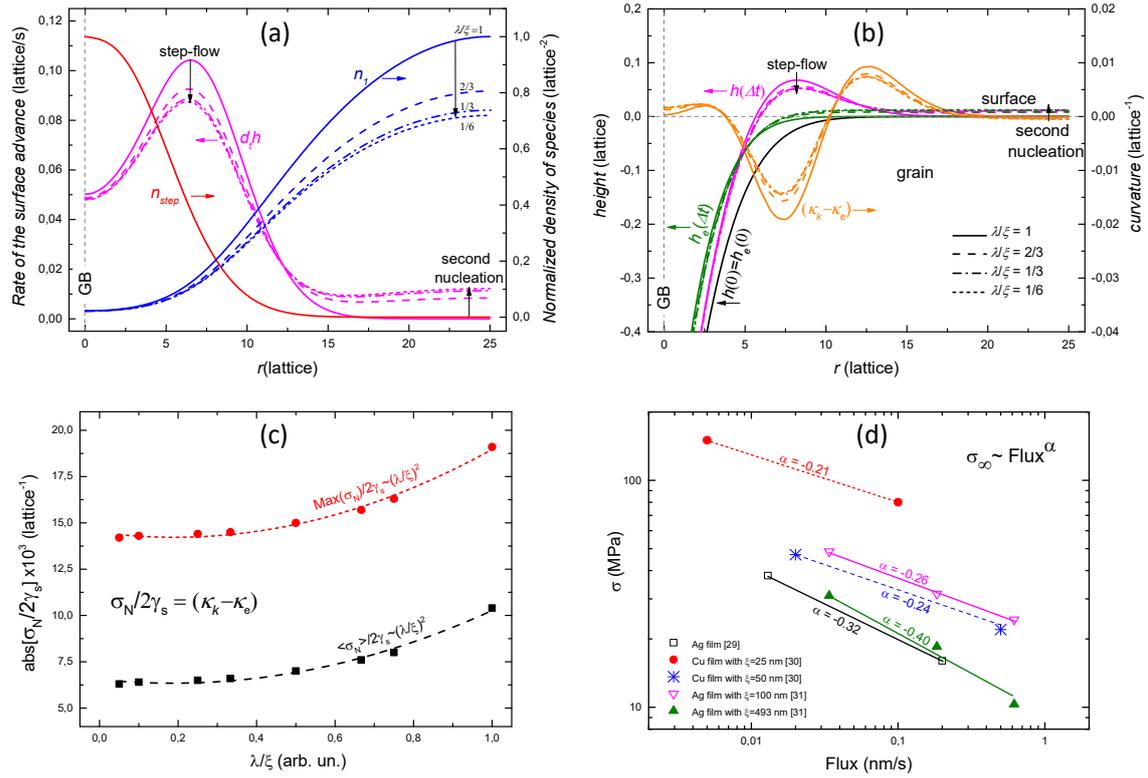

**Fig. 5.** $\lambda/\xi$—Dependence of: **(a)** the steady monomer density ($n_1$, blue curves), the density of A/B-undistinguishable (111) steps ($n_{step}$, red curve), and the local rates of surface advance ($\partial_t h$, pink curves). Vertical arrows indicate the direction of decrease in $\lambda/\xi$ [from solid curve to increasingly dashed ones, as labelled in (a) and (b)]. **(b)** Surface profiles resulting from kinetically limited (pink curve for $\beta$=0.01) and kinetic limitation-free (green curve for $\beta$=0.9) growths for the same deposited volume, together the differences between their curvatures [$(\kappa_k - \kappa_e)$, orange curves]. The regions affected by step-flow and second nucleation are indicated in both figures. **(c)** $\lambda/\xi$—Dependences of $max(\sigma_N)/2\gamma_s = min(\kappa_k - \kappa_e)$ and $\langle\sigma_N\rangle/2\gamma_s = \langle\kappa_k - \kappa_e\rangle$ [red and black symbols computed from data in (b)] together their parabolic fits (dashed curves). **(d)** Fits of the data $\sigma_N$ vs. $F$ (where $\sigma_\infty$ corresponds to the steady postcoalescence value of $\sigma_N$ reported in the literature [29,30,31], see legend) to power-laws dependences.

## 3. CONCLUSIONS

We present an atomistic interpretation of the origin of the non-equilibrium morphology that forms at the edges of the grooves where adatoms accumulate during surface diffusion, according to Mullins' theory. This non-equilibrium morphology is the result of kinetically limited growth by step-edge barriers and reversible nucleation/aggregation processes delaying interlayer transport. As it might be expected, the non-equilibrium morphology gives rise to a surface field of intrinsic stress, whose profile and mean values were estimated by means of the Young-Laplace equation. Once the mechanisms, growth modes (step-flow and second nucleation) and kinetic limitations are identified, we addressed the dependence of the intrinsic stress with the deposition parameters (namely, the deposition flux and temperature for different growth stages).

Finally, the atomistic interpretation presented here complements previous mesoscopic models (those of Refs. [5,6], and also Mullins´ theory [11]) providing a comprehensive and multiscale understanding of the phenomenon of intrinsic stress generation in polycrystalline films and coatings. The kinetics of intrinsic stress in nanocrystalline films will be approached in forthcoming works.

## ACKNOWLEDGMENTS

This work was supported by the Ministerio de Economía, Industria y Competitividad (Spain) under Project No. FIS2017-82415-R; Ministerio de Ciencia, Innovación y Universidades (Spain) within the framework of UE M-ERA.NET 2018 program, under Project StressLIC (Spanish national subprojects PCI2019-103604 and PCI2019-103594). M.J.R.P. acknowledge support by Comunidad de Madrid under contract PEJ-2019-AI/IND-14228. E.G.M. and C.P. acknowledge financial support from the MICINN, through the "María de Maeztu" Programme for Excellence Units in R&D (CEX2018-000805-M).